\documentclass[11pt,twoside]{article}

\usepackage{fancyhdr}
\usepackage{epsfig}
\usepackage{latexsym}
\usepackage{microtype}
\usepackage{lastpage} 
\usepackage{hyperref}
\usepackage{graphicx}
\usepackage{lipsum}
\usepackage{amsmath}
\usepackage{amssymb}
\usepackage[dvipsnames*,svgnames]{xcolor}
\usepackage{graphicx}
\usepackage{hyperref}
\usepackage{algorithmic}
\usepackage{tikz}
\usetikzlibrary{arrows}
\usetikzlibrary{positioning}
\usetikzlibrary{decorations}
\usetikzlibrary{decorations.pathmorphing}
\usepackage{multirow}
\usepackage{comment}
\usepackage{caption}
\usepackage{array}
\hypersetup{colorlinks=true,allcolors=DarkBlue,linkcolor=black}
\usepackage{lineno,hyperref}
\usepackage{amssymb}
\usepackage{amsthm}
\usepackage{amsmath}
\usepackage{mathrsfs}
\usepackage{fixmath}
\usepackage[blocks]{authblk}
\usepackage{multirow}
\usepackage{array}
\usepackage{comment}
\usepackage{graphicx}
\usepackage{wrapfig}

\newcommand{\trshortyear}{20}
\newcommand{\trpapernumber}{03}
\newcommand{\trmonth}{April}
\newcommand{\tryear}{2020}

\DeclareMathOperator*{\argmax}{arg\,max}

\def\R{{\mathbb R}}
\def\P{{\mathcal P}}

\def\x{{\mathbf x}}

\def\v{{\mathbf v}}

\newcommand{\TheAuthor}{}
\newcommand{\Author}[1]{\renewcommand{\TheAuthor}{#1}}

\newcommand{\TheTitle}{}
\newcommand{\Title}[1]{\renewcommand{\TheTitle}{#1}}

\Author{E. F. OLARIU, C. FR\u ASINARU, A. A. POLICIUC}
\Title{A B\&B Algorithm for Coalition Structure Generation over Graphs}

\newcommand\blfootnote[1]{%
	\begingroup
	\renewcommand\thefootnote{}\footnote{#1}%
	\addtocounter{footnote}{-1}%
	\endgroup
}

\begin{document}

	\blfootnote{The publisher does not claim any copyright for the technical reports. The author keeps the full copyright for the paper, and is thus free to transfer the copyright to a publisher if the paper is accepted for publication elsewhere. }   	

\parindent=8mm

\noindent {{\bf \scriptsize Faculty of Computer Science, Alexandru Ioan Cuza University Ia\c si}}

\noindent {{\bf \scriptsize Technical Report TR \trshortyear-\trpapernumber, \trmonth ~ \tryear}}
\vskip -3mm
\noindent\rule{10.2cm}{0.4pt}
\vskip -1mm
\noindent

\vspace{1cm}
\begin{center}
{\Large\bf A Branch and Bound Algorithm for Coalition Structure Generation over Graphs}
\end{center}
\vspace{4mm}

\begin{center}
{\large Emanuel Florentin OLARIU, Cristian FR\u ASINARU, Albert Abel POLICIUC}\\
Faculty of Computer Science, Alexandru Ioan Cuza University Ia\c si, \\
General Berthelot 16, 700483 Ias\c i, Romania, \\
Email: {\tt olariu@info.uaic.ro, acf@info.uaic.ro, abel.policiu@info.uaic.ro}
\end{center}
\vspace{3ex}

\date{}

\begin{abstract}
We give a column generation based branch and bound algorithm for coalition structure generation over graphs problem using valuation functions for which this problem is proven to be NP-complete. For a given graph $ G = (V, E) $ and a valuation function $ w : 2^V \to \R $, the problem is to find the most valuable coalition structure (or partition) of $ V $. We consider two cases: first when the value of a coalition is the sum of the weights of its edges which can be positive or negative, second when the value of a coalition takes account of both inter- and intra-coalitional disagreements and agreements, respectively. For both valuations we give experimental results which cover for the first time sets of more than forty agents. 

For another valuation function (coordination) we give only the theoretical considerations in the appendix.
\smallskip


\end{abstract}

\section{Introduction}
\label{sect0}

Coalition structure generation (CSG) is a major problem in artificial intelligence (\cite{voice12}), multi-agent systems (\cite{bistaffa14}, \cite{voice_12}) communication networks, cooperative game theory (\cite{deng94}, \cite{flammini18}, \cite{ueda18}), scheduling (\cite{hoffman93}), economic theory (like combinatorial auctions) etc. Given a set of agents $ V = \{ 1, 2, \ldots, n \} $, and a {\it valuation function} $ v : 2^V \to \R $ assigning a value to any coalition of agents, the problem is to partition the set of agents into pairwise disjoint {\it classes} ({\it coalitions}) such that the sum of their values is maximized.  CSG is one of the steps in the coalition formation process \cite{sandholm09} which may include also optimizing the coalitions performance and rewarding the coalitions value among the members.

CSG comes from real-world applications: consider a set of agents who can cooperate by working in coalitions. Some of them work better together while others find difficult to cooperate. The problem is to maximize the so called social well-fare or the total value of the designated coalitions. Classical cooperative game theory usually uses a super-additive valuation function that values better a merged coalition than the sum of of the values of the component coalitions. This leads to the great coalition formation which can be costly to coordinate and/or manipulate. Besides these considerations there are natural constraints on possible coalitions, hence the abstractions like super-additivity are not always appropriate for modeling the coalitions values and the agents should be divided into smaller coalitions.

The problem is computationally difficult: the input specification for all $ 2^{|V|} $ possible coalitions is intractable even for reasonable values of $ |V| $ and the computational difficulty maintains even under quite restrictive assumptions. The literature presents a very large number of approaches to CSG problem like (see \cite{rahwan15}): dynamic programming, meta heuristic methods, branch and bound algorithms based on dividing the searching space, anytime algorithms that maintain a monotonically improving feasible solution etc.

Our setting is that of (mixed) integer linear/quadratic programming and is based on the classical model of the set partitioning problem in a graph. Given a graph $ G = (V, E) $ the vertices are agents and the edges represent connections between agents. We study two types of valuation functions: the {\it edge sum} and the {\it correlation} functions. The value of a coalition by the edge sum valuation function (\cite{deng94}) is the sum of the weights of its edges (suppose we have a weight function defined on the set of edges); this function frequently occurs in communication networks and cooperative game theory. The correlation valuation function (\cite{bansal04}) is defined on the entire coalition structure, although it can be defined on coalitions only, and occurs in the clustering framework. This function takes account of the agreements from inside and the disagreements from outside the structure - an edge being labeled with a plus or a minus depending on whether the involved agents are similar or different.

Our approach is based on solving an integer linear programming problem, which is equivalent with CSG, using a branch and bound algorithm in which the problems in nodes are built by means of column generation method. This approach works as long as the involved sub-problem can be solved. We proved that this can be done for edge sum and correlation valuation functions since the sub-problem becomes a quadratic knapsack problem with forbidden configurations. This variant of quadratic knapsack problem can be solved in practice using the corresponding integer quadratic programming model or the mixed integer linear programming equivalent model.

Our numerical results show that we can approach by this method sets of up to $ 45 $ agents depending on the magnitude of the weight function for edge sum valuation function and up to $ 40 $ agents for the correlation valuation function, using a regular home PC.

The paper is organized as follows: section \ref{sect1} describes the LP model and the column generation framework, section \ref{sect2} is dedicated to the two valuation functions outlining the corresponding sub-problems and their linearizations, section \ref{sect3} describes the branch and bound algorithm, and the last section contains the numerical results and the conclusions. The appendix contains ..

\section{LP model and column generation}
\label{sect1}

Consider $ V = \{ 1, 2, \ldots, n \} $ to be a set of $ n $ agents and $ v : 2^V \to \R $ a valuation function on the power set of $ V $. A {\it coalition structure} is a collection of disjoint exhaustive subsets $ C_1, C_2, \ldots, C_p $, where $ C_i \cap C_j = \varnothing $ for all $ 1 \le i < j \le p $ and $ \displaystyle \bigcup_{i = 1}^p C_i = V $.

The problem of finding a coalition structure of maximum value is equivalent with the set partitioning problem (SPP) (see \cite{hoffman93}):
\begin{align}
&\begin{array}{rcl}
& max & \displaystyle  \sum_{j = 0}^{2^n - 1} v(C_j)x_j 
\end{array}
\label{eqa1}\\
&\begin{array}{rcl}
\displaystyle \sum_{j = 0}^{2^n - 1}a_{ij}x_j& = & 1,  \forall i \in V,
\end{array}
\label{eqa2}\\
&\begin{array}{rcl}
\displaystyle x_j & \in & \{ 0, 1 \}, \forall j \in \{ 0, 1, \ldots, 2^n - 1 \}
\end{array}
\label{eqa3}
\end{align}
where $ \{ C_0, C_1, \ldots, C_{2^n - 1} \} $ is an enumeration of $ 2^V $ and $ (a_{ij})_{i \in V} $ is the characteristic vector of $ C_j $, for each $ j $, that is
\begin{align}
a_{ij} = \left\{\begin{array}{rl}
1, & \mbox{if } i \in C_j\\
0, & \mbox{otherwise}
\end{array}  \right., \mbox{ for each } j \in \{ 0, 1, \ldots, 2^n - 1 \}.
\label{eqa4}
\end{align}

We can relax the integrality constraints by replacing \eqref{eqa3}  with
\begin{align}
\tag{3'}
\label{eqa3prim}
\begin{array}{rcl}
\displaystyle x_j & \ge & 0, \forall j \in \{ 0, 1, \ldots, 2^n - 1 \}
\end{array} 
\end{align}

Writing the original relaxed problem as a minimum one means to replace  \eqref{eqa1} by  \eqref{eqa1prim} (by ignoring the minus in front of min)
\begin{align}
\tag{1'}
\label{eqa1prim}
& \begin{array}{rcl}
& min & \displaystyle \left( \sum_{j = 0}^{2^n - 1} -v(C_j)x_j \right)
\end{array}
\end{align}

The dual of the problem \eqref{eqa1prim}, \eqref{eqa2} , \eqref{eqa3prim} is
\begin{align}
%\tag{5}
&\begin{array}{rcl}
& max& \displaystyle \left(\sum_{j = 1}^{n} \pi_i \right)
\end{array}
\label{eqad1}\\
%\tag{6}
&\begin{array}{rcl}
\displaystyle \sum_{i = 1}^n a_{ij}\pi_i & \le &  -v(C_j), \forall j \in \{ 0, 1, \ldots, 2^n - 1 \}
\end{array}
\label{eqad2}\\
%\tag{7}
&\begin{array}{rcl}
\displaystyle \pi_i \in \R, \forall i \in V
\end{array}
\label{eqad3}
\end{align}

Let $ C_0 = \varnothing $, and $ C_j = \{ j \} $, for any $ j \in \{ 1, 2, \ldots, m \} $. An initial feasible basic solution to the problem \eqref{eqa1prim}, \eqref{eqa2}, \eqref{eqa3prim} could be that corresponding to the coalition structure $ \displaystyle \left( \{ 1 \}, \{ 2 \}, \ldots \{m \} \right) $, that is, $ x_1, x_2, \ldots, x_m $. Now we restrict the problem to a small number of variables including $ x_1, x_2, \ldots, x_m $ and we get the restricted master problem (RMP):
\begin{align}
%\tag{1'}
&\begin{array}{rcl}
& min & \displaystyle \left( \sum_{j \in J} -v(C_j)x_j \right)
\end{array}
\label{eqRPM1}\\
%\tag{2'}
&\begin{array}{rcl}
\displaystyle \sum_{j \in J} a_{ij}x_j& = & 1,  \forall i \in V,
\end{array}
\label{eqRPM2}\\
%\tag{3''}
&\begin{array}{rcl}
\displaystyle x_j & \ge & 0, \forall j \in J
\end{array}
\label{eqRPM3}
\end{align}
where $ \displaystyle J  =  \left\{ 1, 2, \ldots, m \right\} \cup J' \subseteq \{ 0, 1, \ldots, 2^m - 1 \} $.

The dual of the RMP is
\begin{align}
%\tag{5}
&\begin{array}{rcl}
& max& \displaystyle \left(\sum_{i = 1}^{n} \pi_i \right)
\end{array}
\label{eqdRPM1}\\
%\tag{6'}
&\begin{array}{rcl}
\displaystyle \sum_{i = 1}^n a_{ij}\pi_i & \le &  -v(C_j), \forall j \in J
\end{array}
\label{eqdRPM2}\\
%\tag{7}
&\begin{array}{rcl}
\displaystyle \pi_i \in \R, \forall 1 \le i \le n
\end{array}
\label{eqdRPM3}
\end{align}

Let $ (\x, \mathbold{\pi}) $ be an optimum primal-dual solution for this pair of problems; we look for a non-basic variable (column) with the minimum negative reduced cost (Dantzig rule) - by solving a corresponding sub-problem - that would be added to the current restricted master problem. When such a variable doesn't exist we can stop: we have an optimum solution to the primal problem. The sub-problem is to find
\begin{align}
\label{eqSubPr2}
& \begin{array}{rcl}
\displaystyle j_0 = \argmax_{j \in \{ 0, 1, \ldots, 2^n - 1 \} \setminus J} \left(v(C_j) + \sum_{i = 1}^na_{ij}\pi_i \right)
\end{array} 
\end{align}

The arising question is: how can we solve problem \eqref{eqSubPr2}? The answer depends on the form of the valuation function $ v(\cdot) $. In the following sections we will analyze this sub-problem for two of the most frequent used valuation functions defined for coalition structure over graphs: the {\it edge-sum} and the {\it correlation} valuation functions.

%The question that arises is: can we solve problem \eqref{eqSubPr2}? The answer depends on the form of the valuation function $ v(\cdot) $. In the following sections we will analyse this sub-problem for three of the most frequent used valuation functions defined for coalition structure over graphs: the {\it edge-sum},  the {\it correlation}, and the {\it coordination} valuation functions.

\section{Coalition Structure Generation over Graphs}
\label{sect2}

\subsection{The edge-sum valuation}
\label{ssES}

Let $ G = (V, E) $ be graph and $ w : E \to \R $ a weight function on its edges. The corresponding {\it edge-sum coalition valuation function} is
\[ v: \P(V) \to \R, v(C) = \sum_{ij \in E : i, j \in C} w_{ij}, \forall C \subseteq V. \] 
This function was extensively studied in the context of cooperative game theory (\cite{deng94}) and the corresponding CSG problem is NP-hard being as hard as the MAX-CUT problem.

Now, if $ \v $ is the characteristic vector of the generic coalition $ C $, then the sub-problem becomes:
\begin{align}
\label{eqSubPr3}
\max_{\v \in \{0, 1 \}^m, v \not= \chi_{C_j}, \forall j \in J}{\left( \sum_{ij \in E} w_{ij} v_i v_j + \sum_{i \in V} \pi_iv_i \right)  } > 0
\end{align}
That is, the sub-problem is a quadratic knapsack problem with forbidden configurations (QKPf). We can find a new column to add to the restricted master problem if and only if this quadratic knapsack problem has a strictly positive optimal objective value.

QKPf could be a computationally difficult problem since the quadratic knapsack problem (QKP) is known to be NP-hard (being a generalization of {\it Clique} problem). By rephrasing it we get:
\begin{align}
& \begin{array}{rcl}
& max& \displaystyle\left( \sum_{ij \in E} w_{ij} v_i v_j + \sum_{i \in V} \pi_iv_i \right)
\end{array}
\label{eqQKPf1}\\
& \begin{array}{rcl}
\displaystyle \sum_{i \in C_j} v_i + \sum_{i \notin C_j} (1 - v_i) & \ge &  1, \forall j \in J
\end{array}
\label{eqQKPf2}\\
&\begin{array}{rcl}
\displaystyle v_i & \in & \{0, 1\}, \forall i \in V
\end{array}
\label{eqQKPf3}
\end{align}

This problem has a quadratic objective but only linear constraints: an Integer Quadratic Programming (IQP) problem. Such a problem can be solved by using a mathematical optimization solver as such or by linearizing it first in order to transform it into a Mixed Integer Linear Programming (MILP) problem. We note here that the known methods used for solving such a problem are not applicable here since the usual requirements are to have natural or positive coefficients for the quadratic (some times non-diagonal) terms (\cite{gallo80}, \cite{pisinger07}). Hence we had to settle to solve it as a general IQP problem using a mathematical optimization solver.

\subsection{The correlation valuation}
\label{ssCorr}

Suppose we have a function $ e : E \to \{ +, - \} $ that assigns to each edge $ ij \in E $ the labels $ + $ or $ - $ depending on whether agents $ i $ and $ j $ are similar or different (this function arises in clustering frameworks). A valuation function that takes account of both inter- and intra-coalitional similarities can be defined in the following way (\cite{bansal04}): we want to minimize the number of mistakes: a positive mistake occurs when $ e(ij) = - $, with $ i $ and $ j $ belonging to the same coalition, a negative one occurs when $ e(ij) = + $, but $ i $ and $ j $ belong to different coalitions. Or, equivalently, we maximize the number of agreements; first define, for a coalition $ C \subseteq V $, the intra- and inter-coalitional connections
\[ Intra^+(C) = | \{ ij \in E \: : \: e(ij) =+, i, j \in C\} |, \]
\[ Inter^-(C) = | \{  ij \in E \: : \: e(ij) =-, i \in C, j \notin C \}  \} |. \] 
The {\it correlation valuation function} is 
\[ v: \P(V) \to \R, v(C) =Intra^+(C) + Inter^-(C)/2, \forall C \subseteq V. \] 
%no edge means no mistake, hence it is better for $ G $ to be a complete graph
It was proven in (\cite{bansal04}) that the corresponding sub-problem is $ NP $-complete. We define two weight functions on the set of edges of $ G $, $ w^+, w^- : E \to \{ 0, 1 \} $, by
\[ w^+(ij) = w^+_{ij} = \left\{ \begin{array}{rl}
1, & \mbox{if } e(ij) = +\\
0, & \mbox{if } e(ij) = -
\end{array} \right., \]
\[  w^-(ij) = w^-_{ij} = \left\{ \begin{array}{rl}
1, & \mbox{if } e(ij) = -\\
0, & \mbox{if } e(ij) = +
\end{array} \right., \forall ij \in E. \]
%$ w^-_{ij} $ is $ 1/2 $ because the coalitions containing $ i $ and $ j $, respectively, must halve the disagreement between $ i $ and $ j $. 

Let $ \v $ be the characteristic vector of the generic coalition $ C $, then
\begin{align}
\notag
& \begin{array}{l}
\displaystyle v(C) =  \frac{1}{2}\sum_{i \in V}\sum_{j \in V}  w^+_{ij}v_iv_j + \frac{1}{2}\sum_{i \in V}\sum_{j \in V} w^-_{ij} v_i(1 - v_j).
\end{array}
\end{align}

The sub-problem \eqref{eqSubPr2} becomes the following QKPf problem:
\begin{align}
& \begin{array}{rcl}
& max& \displaystyle\left( \frac{1}{2}\sum_{i \in V}\sum_{j \in V} w^+_{ij}v_iv_j + \frac{1}{2} \sum_{i \in V}\sum_{j \in V} w^-_{ij} v_i(1 - v_j) + \sum_{i \in V} \pi_iv_i \right)
\end{array}
\label{eqCorrPf1}\\
& \begin{array}{rcl}
\displaystyle \sum_{i \in C_j} v_i + \sum_{i \notin C_j} (1 - v_i) & \ge &  1, \forall j \in J
\end{array}
\label{eqCorrPf2}\\
&\begin{array}{rcl}
\displaystyle v_i & \in & \{0, 1\}, \forall i \in V
\end{array}
\label{eqCorrPf3}
\end{align}

\subsection{The linearization of QKPf}
\label{sssLinES}

We present here the linearization due to Glover (\cite{glover75}) for solving QKP. First, we consider the following equivalent form of the objective function quadratic fragment $ \displaystyle \sum_{i = 1}^n \sum_{j = 1}^n w'_{ij}v_iv_j $, where $ 2w'_{ij} = w_{ij} $, for each $ i \not= j $, and $ w'_{ii} = 0 $, for all $ i \in V $. Second,$ \displaystyle \sum_{j = 1}^nw'_{ij} v_iv_j  $ is replaced by a new real variable $ u_i $, and we add four new constraints 
\begin{align}
\notag
& \begin{array}{l}
W_i^{'-} v_i\le u_i \le W_i^{'+} v_i,\\
\displaystyle \sum_{j = 1}^n w'_{ij} v_j - W_i^{'+}(1 - v_i) \le u_i \le \displaystyle \sum_{j = 1}^n w'_{ij} v_j - W_i^{'-}(1 - v_i),
\end{array}
\end{align}
where $ W_i^{'-} $ and $ W_i^{'+} $ are a lower and, respectively, an upper bound for $ \displaystyle \sum_{i = 1}^n \sum_{j = 1}^n w'_{ij} v_j $, e.g.:
\begin{align}
%\notag
\label{eqLoUpB}
& \begin{array}{l}
W_i^{'-} =\displaystyle  \sum_{j: w'_{ij} < 0} w'_{ij} \mbox{ and } W_i^{'+} = \displaystyle \sum_{j: w'_{ij} > 0} w'_{ij}
\end{array}
\end{align}

The problem \eqref{eqQKPf1} - \eqref{eqQKPf3} becomes, after replacing $ w'_{ij} $, $ W_i^{'-} $, and $ W_i^{'+} $, by correspondingly $ w_{ij} $, $ W_i^{-} $, and $ W_i^{+} $ (and substituting $ 2u_i $ by $ u_i $)
\begin{align}
& \begin{array}{rcl}
& max& \displaystyle \left( \sum_{i \in V} \pi_iv_i + \sum_{i \in V} u_i \right)
\end{array}
%\label{eqGlover1}\\
\notag\\
& \begin{array}{rcl}
\displaystyle \sum_{i \in C_j} v_i + \sum_{i \notin C_j} (1 - v_i) & \ge &  1, \forall j \in J,
\end{array}
\notag\\
%\label{eqGlover2}\\
& \begin{array}{rcl}
\displaystyle \sum_{j: ij \in E} w_{ij} v_j + W_i^{+} v_i - u_i & \le & W_i^{+}, \forall i \in V,
\end{array}
\notag\\
%\label{eqGlover3}\\
& \begin{array}{rcl}
\displaystyle \sum_{j: ij \in E} w_{ij} v_j + W_i^{-} v_i - u_i & \ge & W_i^{-}, \forall i \in V,
\end{array} 
\notag\\
%\label{eqGlover4}\\
& \begin{array}{rcl}
W_i^{-} v_i - u_i \le 0, \forall i \in V,
\end{array} 
\notag\\
%\label{eqGlover5}\\
& \begin{array}{rcl}	
 W_i^{+} v_i - u_i  \ge  0, \forall i \in V,
\end{array} 
\notag\\
%\label{eqGlover6}\\
& \begin{array}{rcl}
\displaystyle v_i & \in & \{ 0, 1 \}, \forall i \in V,
\end{array} 
\notag\\
%\label{eqGlover7}\\
& \begin{array}{rcl}
\displaystyle u_i & \in & \R, \forall i \in V.
\end{array} 
\notag
%\label{eqGlover8}
\end{align}

For the correlation valuation function the corresponding model can be linearized as in a similar way; replacing the coefficients of the linear fragment with $ \displaystyle p_i = \frac{1}{2} \sum_{j \in V} w^-_{ij} + \pi_i, \forall i \in V $, we get the following linear formulation of the problem \eqref{eqCorrPf1} - \eqref{eqCorrPf3}:
\begin{align}
& \begin{array}{rcl}
& max& \displaystyle \left( \sum_{i \in V} p_iv_i + \sum_{i \in V} u_i \right)
\end{array}
\notag\\
%\label{eqGlover1}\\
& \begin{array}{rcl}
\displaystyle \sum_{i \in C_j} v_i + \sum_{i \notin C_j} (1 - v_i) & \ge &  1, \forall j \in J,
\end{array}
\notag\\
%\label{eqGlover2}\\
& \begin{array}{rcl}
\displaystyle \sum_{j: ij \in E} w'_{ij} v_j + W_i^{'+} v_i - u_i & \le & W_i^{'+}, \forall i \in V,
\end{array}
\notag\\
%\label{eqGlover3}\\
& \begin{array}{rcl}
\displaystyle \sum_{j: ij \in E} w'_{ij} v_j + W_i^{'-} v_i - u_i & \ge & W_i^{'-}, \forall i \in V,
\end{array} 
\notag\\
%\label{eqGlover4}\\
& \begin{array}{rcl}
W_i^{'-} v_i - u_i \le 0, \forall i \in V,
\end{array} 
\notag\\
%\label{eqGlover5}\\
& \begin{array}{rcl}	
 W_i^{'+} v_i - u_i \ge 0, \forall i \in V,
\end{array} 
\notag
%\label{eqGlover6}\\
\end{align}
\begin{align}
& \begin{array}{rcl}
\displaystyle v_i & \in & \{ 0, 1 \}, \forall i \in V,
\end{array} 
\notag\\
%\label{eqGlover7}\\
& \begin{array}{rcl}
\displaystyle u_i & \in & \R, \forall i \in V,
\end{array} 
\notag
%\label{eqGlover8}
\end{align}
where $ w'_{ij} = 0.5 \cdot (w^+_{ij } - w^-_{ij }) $, for all $ i, j \in V $, and the upper and lower bounds ($ W_i^{'+} $ and $ W_i^{'-} $) may have the same expressions like in \eqref{eqLoUpB}.

These last problems are Mixed Integer Linear Programming (MILP) problems that have a reasonable size and can theoretically be solved by the means of a mathematical optimization solver. We presented these linear models for the sake of completeness, but the numerical results for them are, by now, - in terms of running time performance - not encouraging. The optimization solver we used performs better on the quadratic models.

\section{Branch and Bound algorithm}
\label{sect3}

Suppose that the current node in the branching tree has a subset of already covered agents, $ \displaystyle U = \displaystyle \bigcup_{j \in J: x_j = 1} C_j $, and $ U' = V \setminus U $. For the edge-sum valuation the sub-problem becomes:
\begin{align}
& \begin{array}{rcl}
& max & \displaystyle \left( \sum_{i \in U'} \sum_{j \in U'} w_{ij}v_i v_j +\sum_{i \in U'} \pi_iv_i \right)
\end{array}
\label{eqBBESPf1}\\
& \begin{array}{rcl}
\displaystyle \sum_{i \in C_j} v_i + \sum_{i \notin C_j} (1 - v_i) & \ge &  1, \forall j \in J, \mbox{ s. t. } C_j \subseteq U'.
\end{array}
\label{eqBBESPf2}\\
&\begin{array}{rcl}
\displaystyle v_i & \in & \{0, 1\}, \forall i \in U'.
\end{array}
\label{eqBBESPf3}
\end{align}

For the correlation valuation function the sub-problem \eqref{eqSubPr2} becomes:
\begin{align}
& \begin{array}{rcl}
& max& \displaystyle\left[ \frac{1}{2}\sum_{i \in U'} \sum_{j \in U'} (w^+_{ij} - w_{ij}^-)v_i v_j +\sum_{i \in U'} \left( \frac{1}{2} \sum_{j \in V} w^-_{ij}+  \pi_i \right)v_i \right]
\end{array}
\label{eqBBCorrPf1}\\
& \begin{array}{rcl}
\displaystyle \sum_{i \in C_j} v_i + \sum_{i \notin C_j} (1 - v_i) & \ge &  1, \forall j \in J, \mbox{ s. t. } C_j \subseteq U'.
\end{array}
\label{eqBBCorrPf2}\\
&\begin{array}{rcl}
\displaystyle v_i & \in & \{0, 1\}, \forall i \in U'.
\end{array}
\label{eqBBCorrPf3}
\end{align}
Both these problems have linearization variants.

At each node of the branching tree we first build the current LP relaxation in two steps: (1) we reduce the variables number by taking account of the branching variables along the path to the root, and (2) we add the necessary variables to the parent node LP relaxation by the means of column generation method. 

Step (1) is implemented by effectively fixing to $ 0 $ the variables $ x_h $ such that $ C_h \cap U \not= \varnothing $ (or, equivalently, $ C_h \cap C_j \not= \varnothing $, for some branching variable $ x_j $ set to $ 1 $), and removing the branching variables $ x_i $ set to 0. The fixing procedure can be achieved by looking in each equation \eqref{eqa2} that contains a branching variable set to $ 1 $. In this way any branching variable set to $ 1 $, that corresponds to a medium sized coalition, has the effect of drastically reducing the size of the corresponding mathematical programming model.

While step (1) is basically the same for both valuation functions, step (2) depends on the specific sub-problem. Step (2) consists in repeatedly solving the corresponding sub-problem while the optimum objective function value is strictly positive. The implementation, however, requires this value to be positive within some tolerance. If this step would not have been subjected to numerical restriction, then the algorithm would have been an exact method.

Our branching rule works in a classical way: we choose a variable, $ x_j $, from the optimal solution in the current node of the branching tree such that its value is around $ 0.5 $ (e. g. $ x_j \in (0.35, 0.65) $ - if possible). It is worth-noting that, for our specific problems, this enumerative method (branch and bound) doesn't need upper bounds because in almost all cases a feasible solution occurs very early - mostly after performing one of the first type (1) steps. The overall effect is that the branching tree has a medium size: we limited the number of branching tree size to $ 40 $, but this bound was hardly reached.

\section{Numerical results}
\label{sect4}

In this section we evaluate the performance of our algorithm. A major part of the running time of our algorithm is concentrated in the root node of the branching tree, where step (2) adds hundreds of new variables to the root model, hence hundreds of sub-problems to solve (fortunately Gurobi solver quickly finds optimal solutions to these QKPf problems), while for the other nodes of the tree the number of such sub-problems drastically reduces.

The algorithm has been written in Java and run on an Intel(R) Core (TM) i5-7500 CPU 3.40GHz computer
with 8GB RAM, under Ubuntu 18.04.4 LTS. The linear and quadratic programming problems were solved using Gurobi 9.0 under an Academic License.

Since there are no benchmarks in the literature for edge-sum or correlation valuation functions, the algorithm has been tested on randomly generated problems. Our test problems were built using the Gilbert model, that is, each edge has a fixed probability of being present in the graph, independently of any other edges (probability $ 1 $ gives a complete graph). The weights on edges are independently generated using a Gaussian distribution $ N(0, 0.2) $ (a certain similar valuation function occurs in \cite{rahwan07}). The name of the benchmark file indicates the probability that an edge belongs to the graph ("p"), the number of agents ("n"), and the number of the instance ("s"). 
%\vspace{-10mm}
\begin{table}[!htbp]
\caption{Solutions and computational times for different instances.}
\label{table1}
\footnotesize
\begin{tabular}{>{\centering}m{0.01\textwidth}>{\centering}m{0.13\textwidth}>{\centering}m{0.05\textwidth}>{\centering}m{0.05\textwidth}>{\centering}m{0.05\textwidth}>{\centering}m{0.05\textwidth}>{\centering}m{0.09\textwidth}>{\centering}m{0.05\textwidth}>{\centering}m{0.08\textwidth}>{\centering}m{0.08\textwidth}c}
\hline
& \multirow{2}{*}{Instance} &  \multicolumn{2}{c}{Solution} & \multicolumn{3}{c}{CPU time (seconds)} & \multicolumn{3}{c}{Number of} & \multirow{2}{*}{Gap}\\
\cline{3-10}& 
& LP & ILP  & overall & root & per node & nodes  & variables & int. sol. &  \\
\hline
\multirow{20}{*}{\rotatebox[origin=c]{90}{Edge-sum valuation}} & p0.8n40s0 & 17.877 & 17.866 & 775 & 547 & 258.3 & 3 & 316 & 1 & 0.06\% \\
&p0.8n40s1 & 21.495 & 21.353 & 1,833 & 1,027 & 122.2 & 15 & 849 & 1 & 0.66\% \\
&p0.8n40s2 & 20.108 & 20.108 & 720 & 720 & 720.0 & 1 & 253 & 1 & 0.00\% \\
& p0.8n40s3 & 19.327 & 19.290 & 812 & 650 & 162.4 & 5 & 320 & 1 & 0.19\% \\
& p0.8n40s4 & 18.166 & 18.099 & 897 & 584 & 179.4 & 5 & 318 & 1 & 0.36\% \\
%& & & & & & & & &   \\
\cline{2-11}
& p0.8n45s0 & 22.606 & 22.411 & 6,672 & 3,565 & 166.8 & {\bf 40} & 1,598 & 5 & 0.86\% \\
& p0.8n45s1 & 22.580 & 22.480 & 4,451 & 2,902 & 404.63 & 11 & 622 & 2 & 0.44\% \\
& p0.8n45s2 & 21.485 & 21.432 & 2,943 & 1,805 & 196.2 & 15 & 799 & 2 & 0.24\% \\
& p0.8n45s3 & 23.884 & 23.884 & 2,815 & 2,815 & 2,815.0 & 1 & 312 & 1 & 0.00\% \\
& p0.8n45s4 & 21.500 & 21.500 & 4,472 & 4,472 & 4,472.0 & 1 & 284 & 1 & 0.00\% \\
%& & & & & & & & &   \\
\cline{2-11}
& p1.0n35s0 & 17.154 & 17.154 & 346 & 346 & 346.0 & 1 & 203 & 1 & 0.00\% \\
& p1.0n35s1 & 16.427 & 16.383 & 446 & 344 & 89.2 & 5 & 273 & 1 & 0.26\% \\
& p1.0n35s2 & 19.406 & 19.241 & 1,651 & 590 & 47.1 & 35 & 1,231 & 2 & 0.85\% \\
& p1.0n35s3 & 15.500 & 15.388 & 630 & 398 & 90.0 & 7 & 334 & 2 & 0.72\% \\
& p1.0n35s4 & 19.064 & 18.987 & 979 & 414 & 108.7 & 9 & 515 & 1 & 0.40\% \\
%& & & & & & & & &   \\
\cline{2-11}
& p1.0n40s0 & 25.286 & 25.286 & 3,860 & 3,860 & 3,860.0 & 1 & 297 & 1 & 0.00\% \\
& p1.0n40s1 & 23.994 & 23.789 & 3,019 & 2,313 & 335.4 & 9 & 534 & 3 & 0.85\% \\
& p1.0n40s2 & 22.401 & 22.401 & 2,376 & 2,376 & 2,376.0 & 1 & 262 & 1 & 0.00\% \\
& p1.0n40s3 & 21.942 & 21.942 & 3,300 & 3,300 & 3,300.0 & 1 & 274 & 1 & 0.00\% \\
& p1.0n40s4 & 18.342 & 18.228 & 3,545 & 1,655 & 154.1 & 23 & 854 & 4 & 0.62\% \\
\hline
\multirow{20}{*}{\rotatebox[origin=c]{90}{Correlation valuation}}& p0.6pS0.6n35s0 & 224.0 & 224 & 256 & 256 & 256.0 & 1 & 433 & 1 & 0.00\% \\
& p0.6pS0.6n35s1 & 243.0 & 243 & 458 & 350 & 50.8 & 9 & 688 & 2 & 0.00\% \\
&p0.6pS0.6n35s2 & 239.0 & 239 & 451 & 451 & 451.0 & 1 & 671 & 1 & 0.00\% \\
& p0.6pS0.6n35s3 & 241.8 & 241 & 711 & 222 & 28.4 & 25 & 1,684 & 1 & 0.33\% \\
& p0.6pS0.6n35s4 & 248.5 & 248 & 465 & 392 & 155 & 3 & 705 & 1 & 0.20\% \\
%& & & & & & & & &   \\
\cline{2-11}
& p0.6pS0.6n40s0 & 292.0 & 292 & 811 & 811 & 811.0 & 1 & 538 & 1 & 0.00\% \\
& p0.6pS0.6n40s1 & 298.0 & 298 & 1,778 & 1,778 & 1,778.0 & 1 & 1,130 & 1 & 0.00\% \\
& p0.6pS0.6n40s2 & 289.0 & 289 & 1,735 & 1,735 & 1,735.0 & 1 & 596 & 1 & 0.00\% \\
& p0.6pS0.6n40s3 & 311.0 & 311 & 938 & 938 & 938.0 & 1 & 500 & 1 & 0.00\% \\
& p0.6pS0.6n40s4 & 313.5 & 313 & 4,542 & 3,062 & 1,514 & 3 & 1,565 & 1 & 0.15\% \\
%& & & & & & & & &   \\
\cline{2-11}
& p0.8pS0.6n35s0 & 304.0 & 304 & 1,714 & 1,277 & 342.8 & 5 & 829 & 3 & 0.00\% \\
& p0.8pS0.6n35s1 & 294.5 & 294 & 1,746 & 834 & 116.4 & 15 & 1,614 & 2 & 0.16\% \\
& p0.8pS0.6n35s2 & 299.0 & 299 & 1,670 & 1,670 & 1,670.0 & 1 & 704 & 1 & 0.00\% \\
& p0.8pS0.6n35s3 & 307.0 & 307 & 1,070 & 1,070 & 1,070.0 & 1 & 567 & 1 & 0.00\% \\
& p0.8pS0.6n35s4 & 298.0 & 298 & 1,137 & 1,137 & 1,137.0 & 1& 422 & 1 & 0.00\%\\
%& & & & & & & & &   \\
\cline{2-11}
& p0.8pS0.6n40s0 & 395.0 & 395 & 4,271 & 4,271 & 4,271.0 & 1 & 839 & 1 & 0.00\% \\
& p0.8pS0.6n40s1 & 380.5 & 380 & 6,643 & 4,522 & 2,214.0 & 3 & 1,500 & 1 & 0.13\% \\
& p0.8pS0.6n40s2 & 385.0 & 385 & 6,402 & 6,402 & 6,402.0 & 1 & 850 & 1 & 0.00\% \\
& p0.8pS0.6n40s3 & 394.0 & 394 & 9,791 & 9,791 & 9,791.0 & 1 & 897 & 1 & 0.00\% \\
& p0.8pS0.6n40s4 & 387.0 & 387 & 7,278 & 7,278 & 7,278 & 1 & 734 & 1 & 0.00\%
\end{tabular}
\end{table}

We slightly generalized the use of the correlation function by supposing that some edges don't have signs at all, that is, allowing an incomplete underlying graph. The random graphs were generated using the same model and an edge has a "$ + $" or a "$ - $" sign with a prescribed probability. The name of a benchmark file for this valuation indicates the probability that an edge belongs to the graph ("p"), the probability of the plus sign ("pSign"), the number of agents ("n"), and finally the number of the instance ("s"). 

Table \ref{table1} shows the results of our numerical tests; it contains the optimum value in the root (LP problem), the (most of the time) optimum value of the ILP problem found by branch and bound algorithm, the overall running time, the running time for finding and solving the LP problem in the root, the running time per node, the number of nodes of the branching tree, the total number of variables added during the execution, the number of nodes fathomed by integrality, and the gap between the LP and the ILP optimum values.

Our algorithm is one of the first to present competitive results for more than $ 20 $ agents for quite general valuation functions, using only a regular PC. For half of the tests the root LP problem has an integer solution and for all of them the gap is less than one percent. The branching trees (with few exceptions) are small and most of the running time is concentrated in the root and in the nodes corresponding to variables set to zero or variables set to one but associated with small sized coalitions. Conducting the branching rule to find medium or large sized coalitions variables seems to be of secondary interest since the resulting trees are small in size. 

We listed only the results using the quadratic models for solving the sub-problem, however the linear models have a good potential which deserves future exploring work (Lagrangian relaxation, special treatment of the high degeneracy etc). Future work should also be directed towards getting rid of the main burden of our algorithm: find specific methods for solving QKPf. Another direction of future research could be a cut generation method (finding inequalities defining facets of a subjacent polyhedral relaxation) that tightens the LP relaxations towards a branch and cut algorithm.

%\bibliography{mybibfile}

\begin{thebibliography}{10}
\expandafter\ifx\csname url\endcsname\relax
  \def\url#1{\texttt{#1}}\fi
\expandafter\ifx\csname urlprefix\endcsname\relax\def\urlprefix{URL }\fi
\expandafter\ifx\csname href\endcsname\relax
  \def\href#1#2{#2} \def\path#1{#1}\fi

\bibitem{bansal04} Bansal, N., Blum, A., Chawla, S.,
\newblock \em{Correlation Clustering},  \em{Machine Learning}, 56, 89--113, 2004.

\bibitem{bistaffa14} Bistaffa, F., Farinelli, A., Cerquides, J., Rodriguez-Aguilar, J., Ramchurn, S. D.,
\newblock \em{Anytime coalition structure generation on synergy graphs},  \em{Proceedings of AAMAS'14},13--20, 2014.

%\bibitem{caprara99} Caprara, A., Pisinger, D., Toth, P.,\newblock \em{Exact Solution of the Quadratic Knapsack Problem},  \em{INFORMS Journal on Computing}, 11 (2), 125--137, 1999.
%\newblock \href{ https://doi.org:10.1287/ijoc.11.2.125} {\path{doi:10.1287/ijoc.11.2.125}}.

\bibitem{deng94} Deng, X., Papadimitriou., C., 
\newblock \em{On the complexity of cooperative solution concepts},  \em{Mathematics of Operations Research}, 19 (2), 257--266, 1994.

\bibitem{flammini18} Flammini, M., Monaco, G., Moscardelli, L., Shalom, M., Zaks, S.,
\newblock \em{Online Coalition Structure Generation in Graph Games},\em{Proceedings of AAMAS'18}, 1353--1361, 2018.

\bibitem{gallo80} Gallo, G., Hammer, P.L., Simeone, B., \newblock \em{Quadratic Knapsack Problems},  \em{Mathematical Programming}, 12 (2), 132--149, 1980.

%\bibitem{glover74} Glover, F, Wolsey, E., \newblock \em{Converting the 0-1 polynomial programming problem to a 0-1 linear program},  \em{Operations Research}, 22 (1), 455--460, 1974.

\bibitem{glover75} Glover, F, Wolsey, E.,
\newblock \em{Improved linear integer programming formulations of nonlinear integer problems},  \em{Amanagement Science}, 22 (4), 180--182, 1975.

\bibitem{hoffman93} Hoffman, K. L.., Padberg, M.,
\newblock \em{Solving Airline Crew Scheduling Problems by Branch-and-Cut}, \em{Management Science}, 6 (39), 657--682, 1993.
%\newblock \href{ https://doi.org/10.1002/net.3230190505} {\path{doi:10.1002/net.3230190505}}.

\bibitem{pisinger07} Pisinger, D.,
\newblock \em{The Quadratic Knapsack Problem - a Survey}, \em{Discrete Applied Mathematics}, 155, 623--648, 2007.

\bibitem{rahwan07} Rahwan, T., Ramchurn, S. D., Dang, V. D., Giovanucci, A., Jennings, N. R.,
\newblock \em{Anytime optimal coalition structure generation}, \em{Proceedings of AAAI'07}, 1184--1190, 2007.

%\bibitem{rahwan09} Rahwan, T., Michalak, T. P., Elkind, E., Wooldridge, M., Jennings, N. R.,\newblock \em{An Exact Algorithm for Coalition Structure Generation and Complete Set Partitioning}, \em{Research Report}, Dept of CS, University of Oxford, CS-RR-13-09, 2009.

\bibitem{rahwan15} Rahwan, T., Michalak, T. P., Wooldridge, M., Jennings, N. R.,
\newblock \em{Coalition structure generation: A survey}, \em{Artificial Intelligence}, 229, 139--174, 2015.
%\newblock Springer Verlag, New York, 1997.
%\newblock \href {https://doi.org/10.1007/s10951-008-0072-x}
%  {\path{doi:10.1007/s10951-008-0072-x}}.

\bibitem{sandholm09} Sandholm, T., Larson, K., Andersson, M., Shehory, O., Tohme, F.,
\newblock \em{Coalition structure generation with worst case guarantees}, \em{Artificial Intelligence}, 1--2, 209--238, 2009.

\bibitem{ueda18} Ueda, S., Iwasaki, A., Conitzer, V., Ohta, N., Sakurai, Y., Yakoo, M.,
\newblock \em{Coalition structure generation in cooperative games with compact representations}, \em{Autonomous Agents and Multi-Agent Systems}, 32, 503--533, 2018.

\bibitem{voice12} Voice, T., Polukarov, M., Jennings, N. R., \newblock \em{Coalition Structure Generation over Graphs}, \em{Journal of Artificial Intelligence Research}, 45, 165--196, 2012.

\bibitem{voice_12} Voice, T., Ramchurn, S., Jennings, N. R., \newblock \em{On coalition formation with sparse synergies}, \em{Proceedings of AAMAS'12}, 45, 223--230, 2012.

\end{thebibliography}

\appendix

\section{Appendix: The coordination valuation}
\label{ssCoord}

Let $ G = (V, E) $ be graph and $ C \subseteq V $ be a coalition, we define 
\begin{align}
\notag
\displaystyle n_i(C) = | \{ jk \in E \: : \: j \in C, k \notin C, ij, ik \in E \}|.
\end{align}

The {\it coordination coalition valuation function} (\cite{voice12}) is
\[ v: 2^V \to \R, v(C) = \sum_{i \in C} n_i(C), \forall C \subseteq V. \] 
This function accounts for all cliques of three agents, two of them being inside the coalition, while the third is outside. It is designed to offer coalitions that include agents that have common neighbours from outside. We will generalize this definition by including the weight of the $ 3 $-clique as the sum of the weights on its edges. Hence, we consider first $ w : E \to \R $ to be an weight on the edges of $ G $.
Second, if $ \v $ is the characteristic vector for the generic coalition $ C $, then 
\begin{align}
\notag
& \begin{array}{l}
\displaystyle v(C) = \sum_{i = 1}^n v_i \sum_{j = 1}^n w_{ij} v_j \sum_{k = 1}^n w_{ik} w_{jk} (1 - v_k).
\end{array}
\end{align}
We get the coordination valuation function by taking $ w $ to be the edge characteristic function, i. e., $ w_{ij}  = 1 $, if $ ij \in E $, and $ 0 $ otherwise. The sub-problem \eqref{eqSubPr2} becomes:
\begin{align}
& \begin{array}{rcl}
& max& \displaystyle\left( \sum_{i = 1}^n \sum_{j = 1}^n \sum_{k = 1}^n w_{ij}w_{ik} w_{jk}  v_i v_j  (1 - v_k) + \sum_{i \in V} \pi_iv_i \right)
\end{array}
\label{eqCubPf1}\\
& \begin{array}{rcl}
\displaystyle \sum_{i \in C_j} v_i + \sum_{i \notin C_j} (1 - v_i) & \ge &  1, \forall j \in J
\end{array}
\label{eqCubPf2}\\
&\begin{array}{rcl}
\displaystyle v_i & \in & \{0, 1\}, \forall i \in V
\end{array}
\label{eqCubPf3}
\end{align}

\subsection{Reducing the degree of the objective function}
\label{sssCoordQL}

Using the same method of Glover from (\cite{glover75}), we will transfom the cubic objective function from \eqref{eqCubPf1} to a quadratic one by adding only $ n $ new unknowns and $ 4n $ new (quadratic) constraints. We introduce first the variables
\begin{align}
\notag
& \begin{array}{l}
\displaystyle z_i = v_i \sum_{j = 1}^n v_j\left(\sum_{k = 1}^n w_{ij} w_{ik} w_{jk} (1 - v_k) \right), \forall i \in V.
\end{array}
\end{align}

Second, add the constraints
\begin{align}
& \begin{array}{l}
Y_i^- v_i\le z_i \le Y_i^+ v_i,\forall i \in V  \\
\displaystyle \sum_{j = 1}^n v_j\left(\sum_{k = 1}^n w_{ij} w_{ik} w_{jk} (1 - v_k) \right) - Y_i^+(1 - v_i) \le z_i, \forall i \in V\\
\displaystyle \sum_{j = 1}^n v_j\left(\sum_{k = 1}^n w_{ij} w_{ik} w_{jk} (1 - v_k) \right) - Y_i^-(1 - v_i) \ge z_i, \forall i \in V,
\end{array}
\end{align}
where $ Y_i^- $ and $ Y_i^+ $ are a lower and, respectively, an upper bound for the expression
\begin{align}
\notag
& \begin{array}{l}
\displaystyle \sum_{j = 1}^n v_j\left(\sum_{k = 1}^n w_{ij} w_{ik} w_{jk} (1 - v_k) \right).
\end{array}
\end{align}

$ Y_i^- $ and $ Y_i^+ $ could be
\begin{align}
& \begin{array}{l}
Y_i^- = \displaystyle \left(\sum_{j, k: w_{ij}w_{ik}w_{jk} < 0} w_{ij} w_{ik} w_{jk} \right)\left/2 \right. \\
Y_i^+ = \displaystyle  \left(\sum_{j, k: w_{ij}w_{ik}w_{jk} > 0} w_{ij} w_{ik} w_{jk} \right)\left/2 \right.
\end{array}
\end{align}

The problem \eqref{eqCubPf1} - \eqref{eqCubPf3} becomes
\begin{align}
& \begin{array}{rcl}
& max& \displaystyle \left( \sum_{i = 1}^n \pi_iv_i + \sum_{i = 1}^n z_i \right)
\end{array}
\label{eqCoordGlover1}\\
& \begin{array}{rcl}
\displaystyle \sum_{i \in C_j} v_i + \sum_{i \notin C_j} (1 - v_i) & \ge &  1, \forall j \in J,
\end{array}
\label{eqCoordGlover2}\\
& \begin{array}{rcl}
\displaystyle \sum_{j = 1}^n v_j\left(\sum_{k = 1}^n w_{ij} w_{ik} w_{jk} (1 - v_k) \right) + Y_i^+v_i - z_i \le Y_i^+, \forall i \in V,
\end{array}
\label{eqCoordGlover3}\\
& \begin{array}{rcl}
\displaystyle \sum_{j = 1}^n v_j\left(\sum_{k = 1}^n w_{ij} w_{ik} w_{jk} (1 - v_k) \right) + Y_i^-v_i - z_i \ge Y_i^-, \forall i \in V,
\end{array} 
\label{eqCoordGlover4}\\
& \begin{array}{rcl}
Y_i^- v_i - z_i \le 0, \forall i \in V,
\end{array} 
\label{eqCoordGlover5}\\
& \begin{array}{rcl}
 Y_i^+ v_i - z_i & \ge & 0, \forall i \in V,
\end{array} 
\label{eqCoordGlover6}\\
& \begin{array}{rcl}
\displaystyle v_i & \in & \{ 0, 1 \}, \forall i \in V,
\end{array} 
\label{eqCoordGlover7}\\
& \begin{array}{rcl}
\displaystyle z_i & \in & \R, \forall i \in V.
\end{array} 
\label{eqCoordGlover8}
\end{align}

(For the coordination valuation function we can simply take $ Y_i^- = 0 $ and $ Y_i^+ = n^2 $, for all $ i \in V $.)

This last problem is a Mixed Integer Quadratically Constrained Programming (MIQCP) problem that could also be solved using a mathematical optimization solver. Another way of addressing this problem is to apply again the method of Glover to the expressions
\begin{align}
\notag
& \begin{array}{l}
z_{ij} = \displaystyle v_j\left(\sum_{k = 1}^n w_{ij} w_{ik} w_{jk} (1 - v_k) \right),
\end{array}
\end{align}
which will add $ n^2 $ new variables and $ 4n^2 $ new constraints to the problem.

\subsection{Branch and Bound version for the coordination valuation function}
\label{ssBBCoord}

Suppose as above that the current node in the branching tree has a subset of already covered agents, $ \displaystyle U = \displaystyle \bigcup_{j \in J: x_j = 1} C_j $, and $ U' = V \setminus U $. For this valuation function the sub-problem \eqref{eqSubPr2} becomes:
\begin{align}
& \begin{array}{rcl}
& max& \displaystyle\left[ \sum_{i \in U'} \sum_{j \in U'} \left(\sum_{k \in U'} w_{ij}w_{ik} w_{jk}  v_i v_j  (1 - v_k) + \sum_{k \in U} w_{ij}w_{ik} w_{jk}  v_i v_j \right)  + \right.\\
& &\displaystyle  \left.+ \sum_{i \in U'} \pi_iv_i \right]
\end{array}
\label{eqBBCubPf1}\\
& \begin{array}{rcl}
\displaystyle \sum_{i \in C_j} v_i + \sum_{i \notin C_j} (1 - v_i) & \ge &  1, \forall j \in J,
\end{array}
\label{eqBBCubPf2}\\
&\begin{array}{rcl}
\displaystyle v_i & \in & \{0, 1\}, \forall i \in U'.
\end{array}
\label{eqBBCubPf3}
\end{align}

Using the above degree reduction procedure for the objective function we introduce the variables
\begin{align}
\notag
& \begin{array}{l}
\displaystyle z_i = v_i \sum_{j \in U'} v_j\left(\sum_{k \in U'} w_{ij} w_{ik} w_{jk} (1 - v_k) + \sum_{k \in U} w_{ij}w_{ik} w_{jk} \right), \forall i \in U'.
\end{array}
\end{align}

Second, add the constraints
\begin{align}
\notag
& \begin{array}{l}
Y_i^- v_i\le z_i \le Y_i^+ v_i,\forall i \in V  \\
\displaystyle \sum_{j \in U'} v_j\left(\sum_{k \in U'} w_{ij} w_{ik} w_{jk} (1 - v_k) + \sum_{k \in U} w_{ij}w_{ik} w_{jk} \right) - Y_i^+(1 - v_i) \le z_i, \forall i \in U'\\
\displaystyle \sum_{j \in U'} v_j\left(\sum_{k \in U'} w_{ij} w_{ik} w_{jk} (1 - v_k) + \sum_{k \in U} w_{ij}w_{ik} w_{jk} \right) - Y_i^-(1 - v_i) \ge z_i, \forall i \in U',
\end{array}
\end{align}
where $ Y_i^- $ and $ Y_i^+ $ are a lower and, respectively, an upper bound for the expression
\begin{align}
\notag
& \begin{array}{l}
\displaystyle \sum_{j \in U'} v_j\left(\sum_{k \in U'} w_{ij} w_{ik} w_{jk} (1 - v_k) + \sum_{k \in U} w_{ij}w_{ik} w_{jk} \right).
\end{array}
\end{align}

$ Y_i^- $ and $ Y_i^+ $ could be
\begin{align}
\notag
& \begin{array}{l}
Y_i^- = \displaystyle \left(\sum_{j, k \in U': w_{ij}w_{ik}w_{jk} < 0} w_{ij} w_{ik} w_{jk} \right)\left/2 \right. + \sum_{j \in U', k \in U: w_{ij}w_{ik}w_{jk} < 0} w_{ij}w_{ik} w_{jk}\\
Y_i^+ = \displaystyle  \left(\sum_{j, k \in U': w_{ij}w_{ik}w_{jk} > 0} w_{ij} w_{ik} w_{jk} \right)\left/2 \right. + \sum_{j \in U', k \in U: w_{ij}w_{ik}w_{jk} > 0} w_{ij}w_{ik} w_{jk}
\end{array}
\end{align}

The problem \eqref{eqCubPf1} - \eqref{eqCubPf3} becomes
\begin{align}
& \begin{array}{rcl}
& max& \displaystyle \left( \sum_{i \in U'} z_i  + \sum_{i \in W'} \pi_iv_i \right)
\end{array}
\label{eqBBCoordGlover1}\\
& \begin{array}{rcl}
\displaystyle \sum_{i \in C_j} v_i + \sum_{i \notin C_j} (1 - v_i) & \ge &  1, \forall j \in J,
\end{array}
\label{eqBBCoordGlover2}\\
& \begin{array}{rcl}
\displaystyle \sum_{j \in U'} v_j\left(\sum_{k \in U'} w_{ij} w_{ik} w_{jk} (1 - v_k) + \sum_{k \in U} w_{ij}w_{ik} w_{jk} \right) + Y_i^+v_i - z_i \le Y_i^+, \forall i \in U',
\end{array}
\label{eqBBCoordGlover3}\\
& \begin{array}{rcl}
\displaystyle \sum_{j \in U'} v_j\left(\sum_{k \in U'} w_{ij} w_{ik} w_{jk} (1 - v_k) + \sum_{k \in U} w_{ij}w_{ik} w_{jk}\right) + Y_i^-v_i - z_i \ge Y_i^-, \forall i \in U',
\end{array} 
\label{eqBBCoordGlover4}\\
& \begin{array}{rcl}
Y_i^- v_i - z_i \le 0, \forall i \in U',
\end{array} 
\label{eqBBCoordGlover5}\\
& \begin{array}{rcl}
 Y_i^+ v_i - z_i & \ge & 0, \forall i \in U',
\end{array} 
\label{eqBBCoordGlover6}
\end{align}
\begin{align}
& \begin{array}{rcl}
\displaystyle v_i & \in & \{ 0, 1 \}, \forall i \in U',
\end{array} 
\label{eqBBCoordGlover7}\\
& \begin{array}{rcl}
\displaystyle z_i & \in & \R, \forall i \in U'.
\end{array} 
\label{eqBBCoordGlover8}
\end{align}

(For the coordination valuation function we can simply take $ Y_i^- = 0 $ and $ Y_i^+ = |U'|^2/2 + |U'| \cdot |U| \le 3n^2/4 $, for all $ i \in U' $.)

\end{document}